\documentclass{osa-article}

\journal{osac}
\articletype{Research Article}
\usepackage{color}
\usepackage{amsmath}
\usepackage{graphicx}
\usepackage{subfigure}
\usepackage{float}
\usepackage{lineno}
\usepackage[utf8]{inputenc}

\begin{document}

\title{Optimizing Ghost Imaging via Analysis and Design of Speckle Patterns}

\author{Xinjian Zhang,\authormark{1,3} Siyuan Song,\authormark{1,3} Xiaoping Ma,\authormark{2} Haonan Zhang,\authormark{1} Lei Gai,\authormark{1} Yongjian Gu,\authormark{1} and Wendong Li\authormark{1*}}

\address{\authormark{1}College of Physics and Optoelectronic Engineering, Ocean University of China, Qingdao 266100, China\\
\authormark{2} College of Mathematical and Physical Sciences, Qingdao University of Science and Technology, Qingdao, China\\
\authormark{3}These authors contributed equally to this work and should be considered co-first authors. }

\email{\authormark{*}Corresponding author:liwd@ouc.edu.cn} %% email address is required

% \homepage{http:...} %% author's URL, if desired

%%%%%%%%%%%%%%%%%%% abstract %%%%%%%%%%%%%%%%
%% [use \begin{abstract*}...\end{abstract*} if exempt from copyright]

\begin{abstract}We study the influence rules of the speckle size of light source on ghost imaging, and propose a new type of speckle patterns to improve the quality of ghost imaging. The results show that the image quality will first increase and then decrease with the increase of the speckle size, and there is an optimal speckle size for a specific object. Moreover, by using the random distribution of speckle positions, a new type of displacement speckle patterns is designed, and the imaging quality is better than that of the random speckle patterns. These results are of great significances for finding the best speckle patterns suitable for detecting targets, which further promotes the practical applications of ghost imaging.
\end{abstract}

%%%%%%%%%%%%%%%%%%%%%%%%%%  body  %%%%%%%%%%%%%%%%%%%%%%%%%%
\section{Introduction}
Based on the principle of second-order correlation in quantum optics, the ghost imaging can reconstruct images of objects through illuminating light with speckle patterns to objects and collecting total intensity of light reflected by objects. In recent years, the ghost imaging has attracted extensive attentions and achieved rapid developments in the fields of optical imaging \cite{bina_backscattering_2013,gong_correlated_2011,li_compressive_2020}  and remote sensing\cite{yang_increasing_2015,gong2016three,erkmen_computational_2012} as it has advantages to resist interference of complex environments and take pictures with high-resolution\cite{tong_preconditioned_2021}.
\par In 1995, T. B. Pittman et al. performed an optical imaging experiment by using photon pairs with quantum entanglement\cite{pittman_optical_1995}. At first, some scholars thought entangled light was necessary for realizing ghost imaging. Later, it was discovered that pseudo-thermal light and real thermal light can also be used to achieve ghost imaging\cite{zhang_correlated_2005,chen_lensless_2009}, which makes the imaging easier to be fulfilled. To simplify the optical systems in ghost imaging, Shapiro proposed the theory of computational ghost imaging(CGI) in 2008 \cite{shapiro_computational_2008}. This scheme uses a spatial light modulator to produce a series of light patterns with randomly distributed speckle (called speckle patterns) as the light source through modulating the phase or intensity of the light, and transforms the optical system of ghost imaging from the original dual-arm structure to a single-arm structure. In 2009, Bromberg et al. prepared the speckle patterns by the spatial light modulator and completed the computational ghost imaging experiment based on the theory of Shapiro \cite{bromberg_ghost_2009}.
\par In the ghost imaging system, the speckle light field are used as pseudo-thermal light sources, and the characteristics of speckle patterns, such as type, speckle size, and quantity, are important factors that determine the imaging quality. Therefore, some researchers have studied the ghost imaging with different speckle light fields \cite{chan_high-order_nodate,__2015,chen_application_2014,wang_effective_2020,zhou_cheng_hybrid_2016,mao_speckle-shifting_2016,nie_noise-robust_2021,li2021sub,zerom_thermal_2012,xuyang_xu_optimization_2015,wang_compressed_2020,cai_hongji_reflection_2019}and made important progresses. Chan et al. \cite{chan_high-order_nodate} theoretically explained the relationship between random speckle patterns and imaging quality, and proposed methods to improve imaging quality. Kong Qingnan et al. \cite{__2015}  generated the speckle patterns by transmitting the laser beam through a rotary ground glass and found that the peak signal-to-noise ratio (PSNR) of the reconstructed image is significantly improved as the average size of speckles increases. In 2014, Mingliang Chen et al. \cite{chen_application_2014} demonstrated the multi-correlation-scale speckle patterns are more efficient and have better antinoise ability than constant-correlation-scale speckle patterns in ghost image. Xiaoxia Wang, Cheng Zhou et al.\cite{wang_effective_2020,zhou_cheng_hybrid_2016} adopted hybrid speckle patterns with different speckle sizes to retrieve the target area with different resolutions and enhance the quality of ghost image. In 2016, Tianyi Mao et al. \cite{mao_speckle-shifting_2016}  introduced the speckle-shifting ghost imaging through shifting the random speckle patterns to improve the performance of edge detection. In 2021, Xiaoyu Nie et al. \cite{nie_noise-robust_2021,li2021sub}proposed a computational ghost imaging scheme using noise speckle patterns illumination, which improved the ability to suppress environmental noise. According to the above research results, it can be seen that the speckle pattern is very important and the speckle size has a direct impact on the quality of ghost imaging. However, the influences of speckle sizes on the imaging quality need to be further revealed in detail. Meanwhile, whether the difference of the position distribution of speckles can affect the ghost imaging has not been considered so far. 
\par In this paper, we will explore the influence rules of the speckle size and the position of the speckle on the imaging quality, and propose a new type of speckle patterns that considers the position distribution of speckles. Firstly, we introduce the random speckle patterns with different speckle sizes, and propose displacement speckle patterns optimally designed by considering the position distribution of speckles. Then we explore the influences of speckle sizes and number of reconstructions on ghost imaging and compare the qualities of ghost image reconstructed by displacement speckle patterns and random speckle patterns. Finally, we demonstrated experimentally the effect of speckle sizes on ghost imaging.
\section{The optimal designs of speckle patterns in the CGI}
The schematic diagram of the CGI is shown in Fig.1. Here, the digital micromirror device (DMD) is used to generate pre-designed speckle patterns, and to project the speckle patterns ${I_{}}(x,y)$ onto the object. Note that, the bucket detector only records the total intensity $ S $ of the light reflected by the object. The object image $ G(x,y) $ can be retrieved by calculating the second-order correlation function between the intensity distribution $ {{I_r}(x,y)} $ illuminated onto the object and the total intensity $ S $ recorded by the bucket detector.
\begin{figure}[!ht]
\centering\includegraphics[width=7cm]{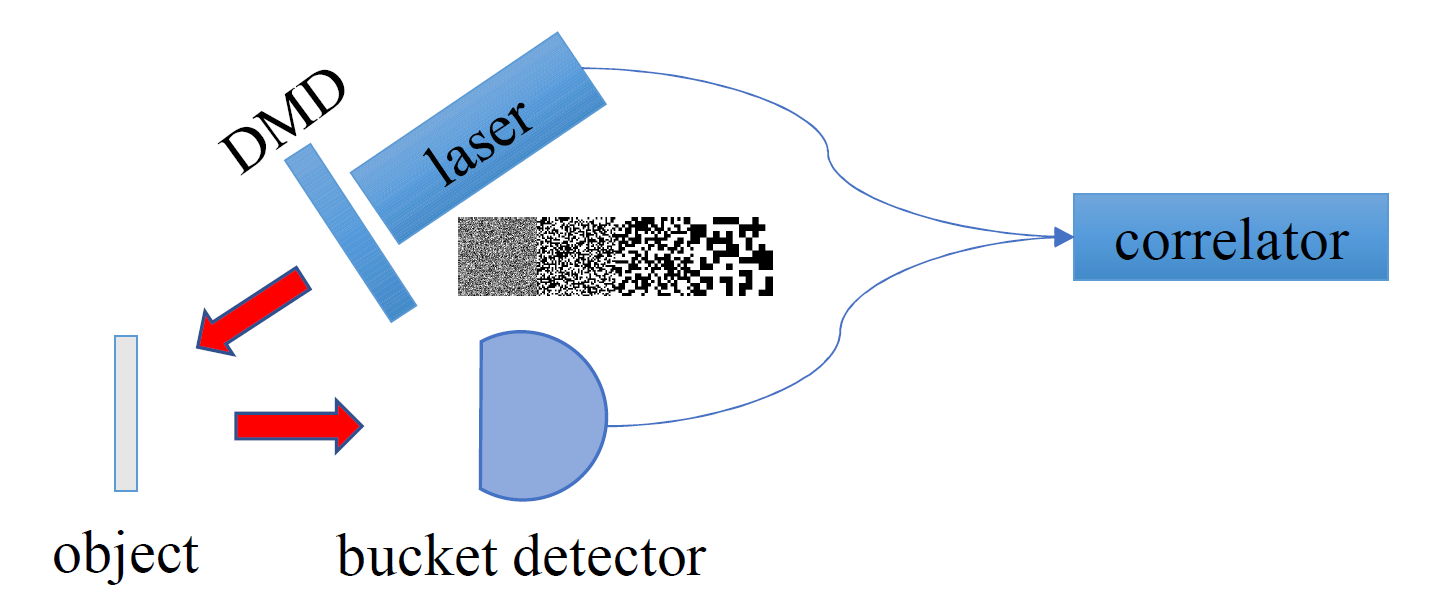}
\caption{The system of computational correlation imaging.}
\end{figure}

\par The second-order correlation function of CGI is:
\begin{equation}
G(x,y) = \frac{1}{N}\sum\limits_{r = 1}^N {(S - \left\langle S \right\rangle ){I_r}(x,y)}
 = \left\langle {{\rm{S}}{I_r}(x,y)} \right\rangle {\rm{ - }}\left\langle {\rm{S}} \right\rangle \left\langle {{I_r}(x,y)} \right\rangle ,
\end{equation}
where  $ \left\langle  \bullet  \right\rangle $ is an average over multiple measurements.
\par In this paper, in order to improve the imaging quality of ghost imaging, we will consider two types of speckle patterns. One is the random speckle patterns, and the other is a new type of optimized displacement speckle patterns proposed below. Here, we define the unit of the speckle size is pixel for convenience. As an example, the random speckle patterns with different speckle sizes are shown in Fig.2.
\begin{figure}[!ht]
\centering\includegraphics[width=7cm]{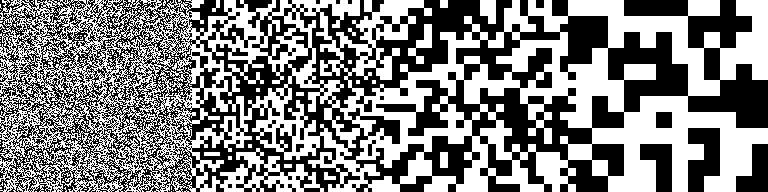}
\caption{The random speckle patterns with different speckle sizes. The speckle sizes from left to right are 1*1, 4*4, 8*8, 16*16, respectively.}
\end{figure}
\par In the CGI, a group of random speckle patterns with the same speckle size is normally used to reconstruct one image. In such cases, the image details are lost seriously when the speckle size is larger, because the size of the minimum lump of gray value in the ghost images is the same with the speckle size, which can be seen from the simulation results in the next section. At the same time, the selections of speckle sizes are greatly restricted as they need to be divisible by the total number of pixels in one row or column of speckle patterns. In order to overcome these shortcomings of the random speckle patterns, we propose a method to generate the speckle patterns with random distribution of speckle positions, which is called displacement speckle patterns.
\par The implementation processes of the displacement speckle patterns include five steps, as follows (suppose the speckle size is n*n (n is an integer greater than 0)):
\par (1) Generate the uniformly distributed random matrix with the dimensions of N*N;
\par (2) Set the threshold to select random matrix elements, and record the coordinates (xi, yi) \par corresponding to the selected matrix elements;
\par (3) Generate an empty martix with the resolution of N*N;
\par (4) Take the coordinates (xi, yi) recorded as the center to select areas with size n*n in the \par empty martix generated in the third step as speckles;
\par (5) Set the color of speckles selected in the fourth step to white, and fill the rest of the  martix\par elements with black.

\par In this way, the displacement speckle patterns can be generated. Their speckle sizes have no restrictions like that in random speckle patterns, and can be any integers smaller than N. As an example, the displacement speckle patterns are shown in Fig.3.
\begin{figure}[!ht]
\centering\includegraphics[width=7cm]{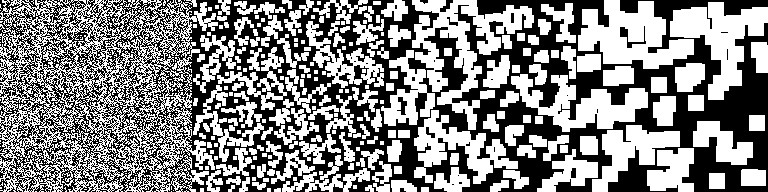}
\caption{The displacement speckle patterns with different speckle sizes. The speckle sizes from left to right are 1*1, 4*4, 8*8, 16*16, respectively.}
\end{figure}
\section{Analyses of the influences of speckle patterns on ghost imaging}
\subsection{Effects of the speckle size and the number of reconstructions on ghost imaging}
\par In this section, with random speckle patterns and displacement speckle patterns, imaging processes of two grayscale objects (the ouc image and the lena image) are numerically simulated, and then compared.
\par In our simultion, we provide 9 types of random speckle patterns, of which the speckle sizes are 1*1, 2*2, 3*3, 4*4, 6*6, 8*8, 12*12, 16*16, 24 *24, respectively, and 24 types of displacement speckle patterns with speckle sizes from 1*1 to 24*24. Note that the number of reconstructions is 80,000 and the resolution is 192*192. Besides, to quantitatively compare the quality of the reconstructed images of objects, we record the PSNR\cite{zhang2014object} every 1000 speckle patterns. The PSNR can be defined by the following expression,
\begin{equation}
\mathrm{PSNR}=10 \log _{10}\left[\frac{\left(2^{\mathrm{m}}-1\right)^{2}}{\mathrm{MSE}}\right],
\end{equation}
where MSE represents the mean square error between the original image and the target image, $m$ is the bit depth of the image which is the number of binary bits used to store each pixel. Typically for grayscale images in the computer, $m$=8.
\par (1) The numerical simulation with random speckle patterns
\par We first use the generated random speckle patterns for numerical simulation. The simulation results of the ouc image and the lena image are shown in Fig.4. The contour plots of the PSNR of the reconstructed images with different speckle sizes and numbers of reconstructions are shown in Fig.5.
\begin{figure}[htbp]
\centering
\subfigure[]{
\includegraphics[width=5.5cm]{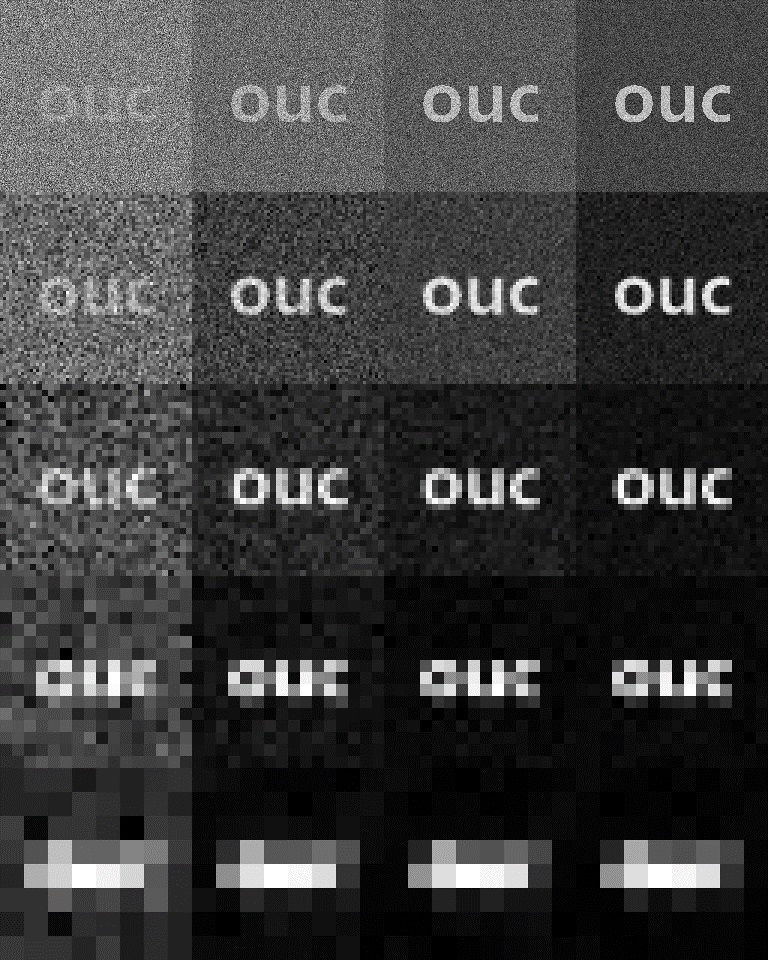}
%\caption{fig1}
}
\quad
\subfigure[]{
\includegraphics[width=5.5cm]{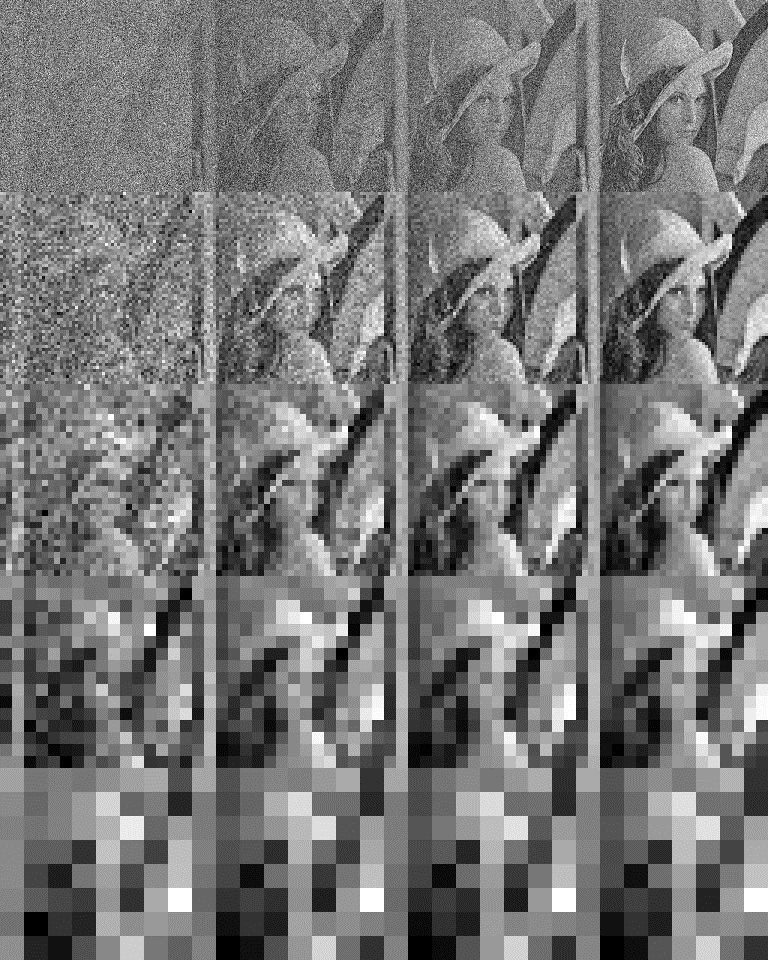}
}
\caption{Ghost images with random speckle patterns: (a) ouc images. (b) lena images. Where the speckle sizes in each row from top to bottom are 1*1, 3*3, 6*6, 12*12, 24*24, the number of reconstructions in each column from left to right is 1000, 10000, 30000, and 80,000, respectively.}
\end{figure}
\par  From Fig.4 and Fig.5, we can find that for the certain number of reconstructions, the image quality and the PSNR perform the rules of first gradually increase and then gradually decrease with the increase of the speckle size. When the number of reconstructions is 1000, the images reconstructed by the speckle patterns with 1*1 speckle size have not only high noise but also low contrast; The PSNR of the image gradually improve with increasing of the speckle size. When the speckle sizes are 10*10 (for the ouc image) and 3*3 (for the lena image), the PSNR reaches the maximum;When the speckle size increase to 24*24, the information of the reconstructed image
\begin{figure}[htbp]
\centering
\subfigure[ouc]{
\includegraphics[width=5.5cm]{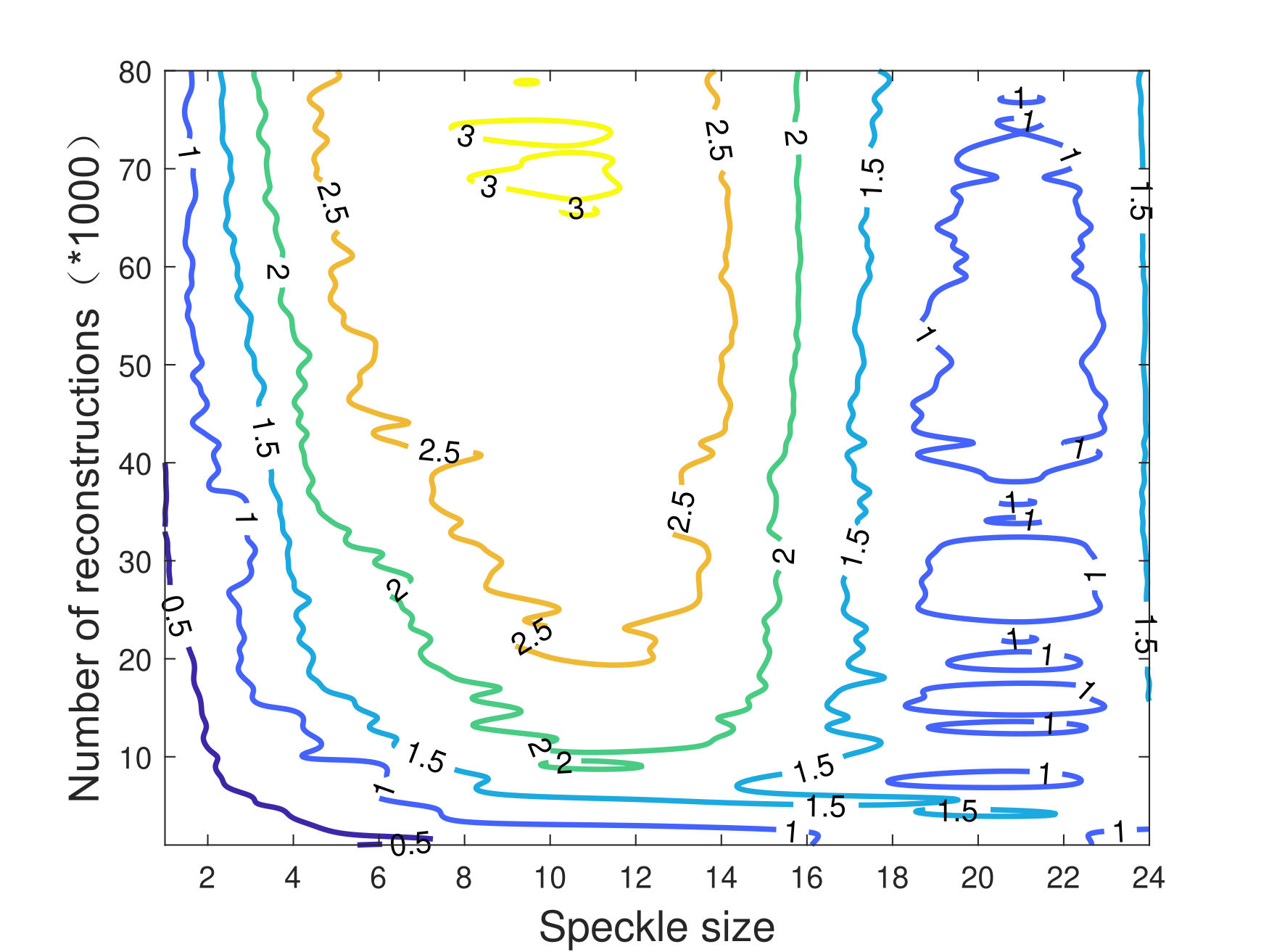}
%\caption{fig1}
}
\quad
\subfigure[lena]{
\includegraphics[width=5.5cm]{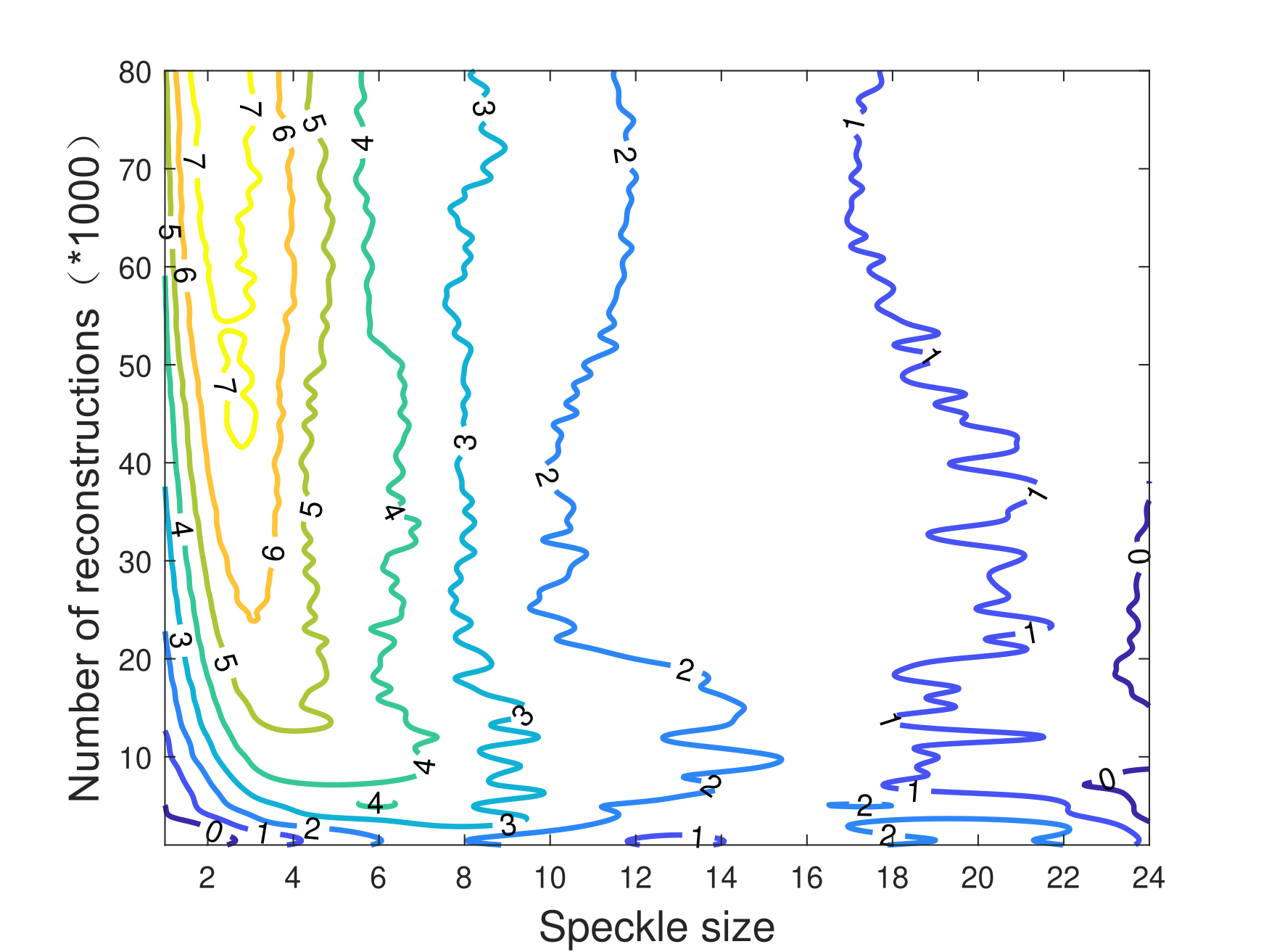}
}
\caption{The contour plots of the PSNR as a function of the speckle size and the number of reconstructions.}
\end{figure}
is severely lost, and only the light and dark information of the corresponding area is retained. If we are keeping the speckle size the same, with the number of reconstructions increases, the imgaing quality and PSNR greatly improve at the beginning, and when it reaches a certain value, it increases slowly and gradually stabilizes; The imaging quality becomes significantly better with the increase of the number of reconstructions when the speckle sizes are 3*3 and 6*6, but when the speckle sizes are 12*12 and 24*24, the imaging quality is difficult to be improved by increasing the number of reconstructions.
\par (2) The numerical simulation with displacement speckle pattern
\par Using the displacement speckle patterns, we get the simulation results as shown in Fig.6. The contour plots of the PSNR of the reconstructed image with the speckle sizes and the number of reconstructions are shown in Fig.7.
\begin{figure}[htbp]
\centering
\subfigure[]{
\includegraphics[width=5.5cm]{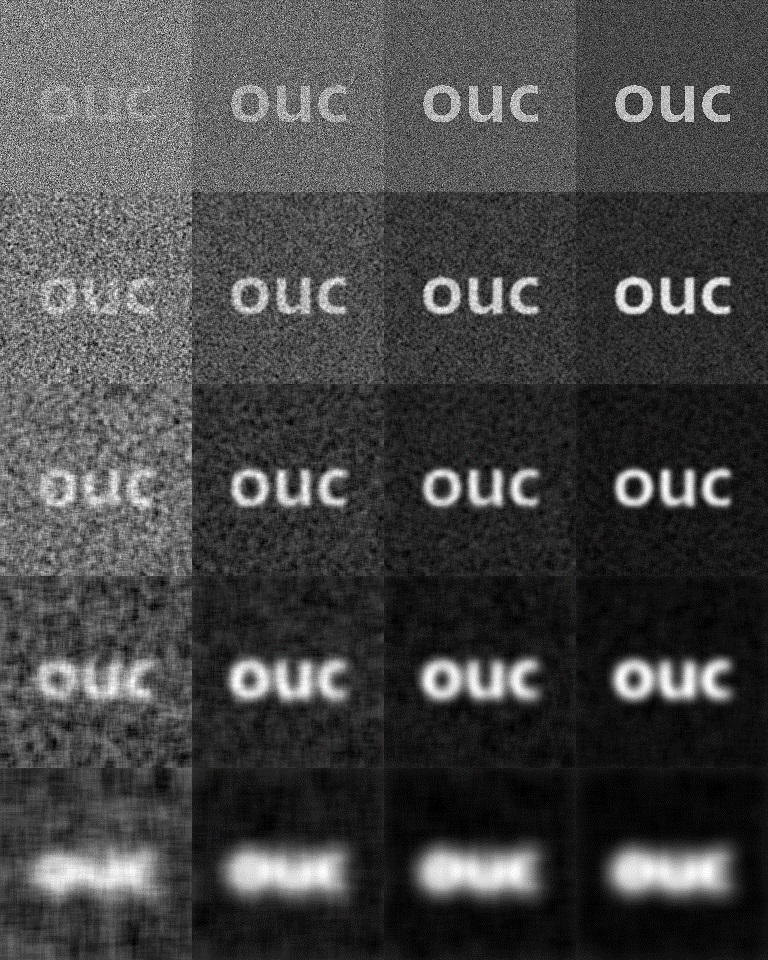}
%\caption{fig1}
}
\quad
\subfigure[]{
\includegraphics[width=5.5cm]{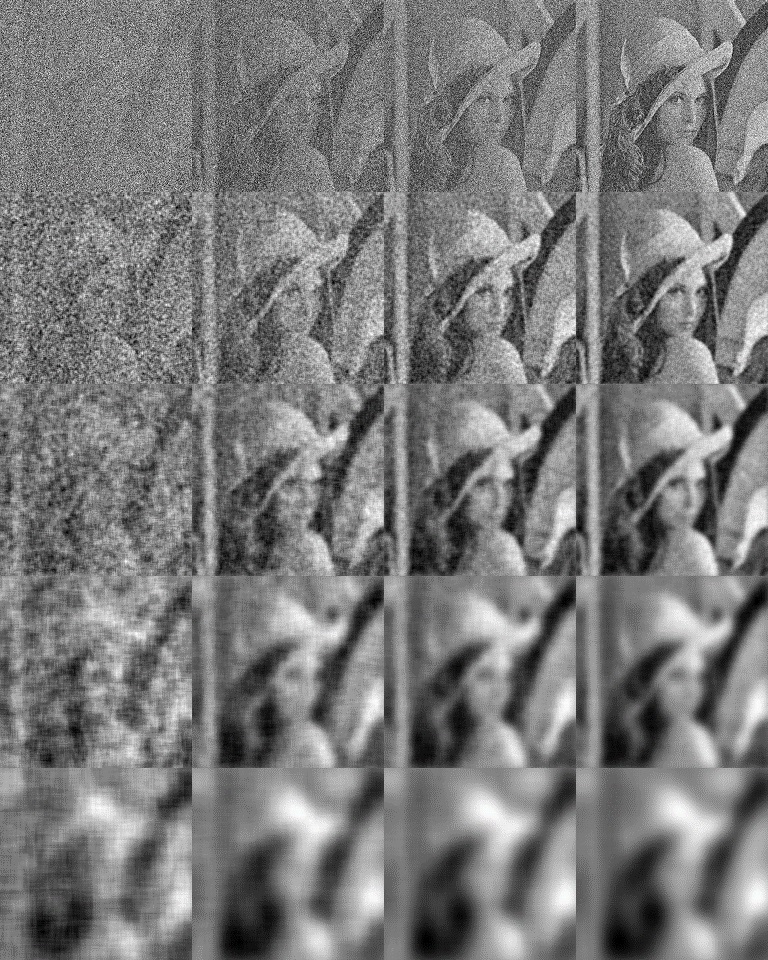}
}
\caption{Ghost images with displacement speckle patterns: (a) ouc images. (b) lena images. Where the speckle size in each row from top to bottom is 1*1, 3*3, 6*6, 12*12, 24*24, the number of reconstructions in each column from left to right is 1000, 10000, 30000, and 80,000, respectively.}
\end{figure}
\begin{figure}[H]
\centering
\subfigure[ouc]{
\includegraphics[width=5.5cm]{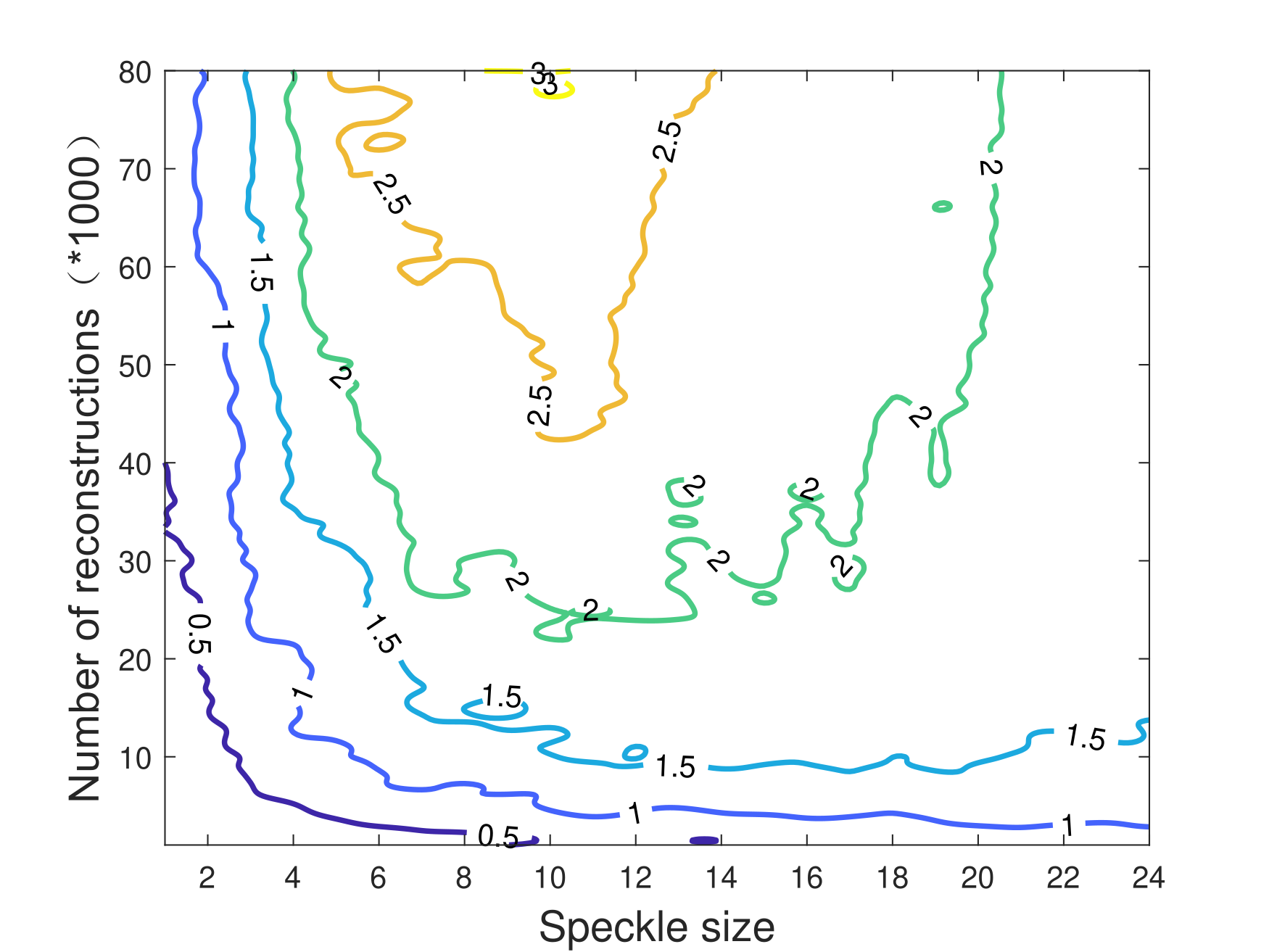}
%\caption{fig1}
}
\quad
\subfigure[lena]{
\includegraphics[width=5.5cm]{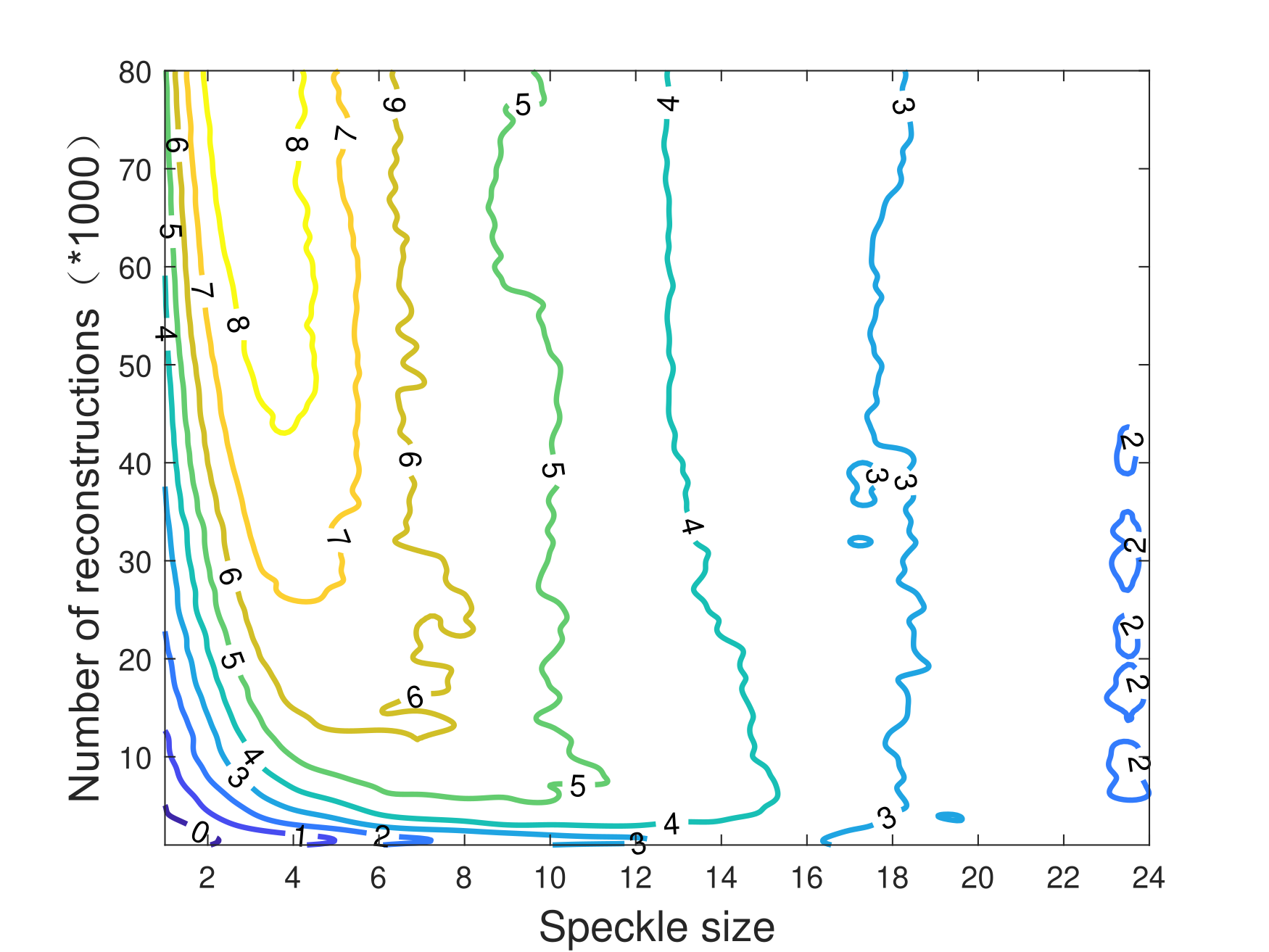}
}
\caption{The contour plots of the PSNR as a function of the speckle size and the number of reconstructions.}
\end{figure}

\par Combined with Fig.6 and Fig.7, it can be seen that whether the ouc iamges with less detailed informations or the informative lena images, the rules of image quality variations with the speckle size and the number of reconstructions under displacement speckle patterns are similar with the results under random speckle patterns. For the comparison between the displacement speckle patterns and the random speckle patterns, we will discuss in the next section.
\par From above analysis, it can be concluded that the quality of the ghost imaging and the PSNR show the rules of first increasing and then decreasing with the increase of the speckle size. For a certain number of reconstructions, there is an optimal speckle size; when the speckle size takes the optimal value, the reconstructed image takes into account the details and the signal-to-noise ratio, which can achieve the best imaging quality. Moreover, for the specific objects, the optimal speckle size decreases with the increase of the number of reconstructions. With the increase of the number of reconstructions, the image quality and the PSNR will be greatly improved at the beginning and gradually stabilize.
\subsection{ Comparisons of random speckle pattern and displacement speckle pattern}
\par As shown in Fig.8 (speckle size is 3*3) and Fig.9 (speckle size is 12*12), we found that when the speckle size is small, the imaging effect of the displacement speckle patterns is not much different than that of the random speckle patterns; when the speckle size increases, the displacement speckle patterns shows great advantage. We can clearly observe that when using the speckle pattern with large speckle size, the results of the displacement speckle patterns can retain the original information of the image, and the light-dark transition is smoother, which is more in line with the subjective perception of the human eye, especially for the image with more details like the lena image. Therefore, the displacement speckle patterns is more suitable for ghost imaging, which has significant meaning for improving the clarity of imaging and reducing the sampling time.
\begin{figure}[H]
\centering\includegraphics[width=7cm]{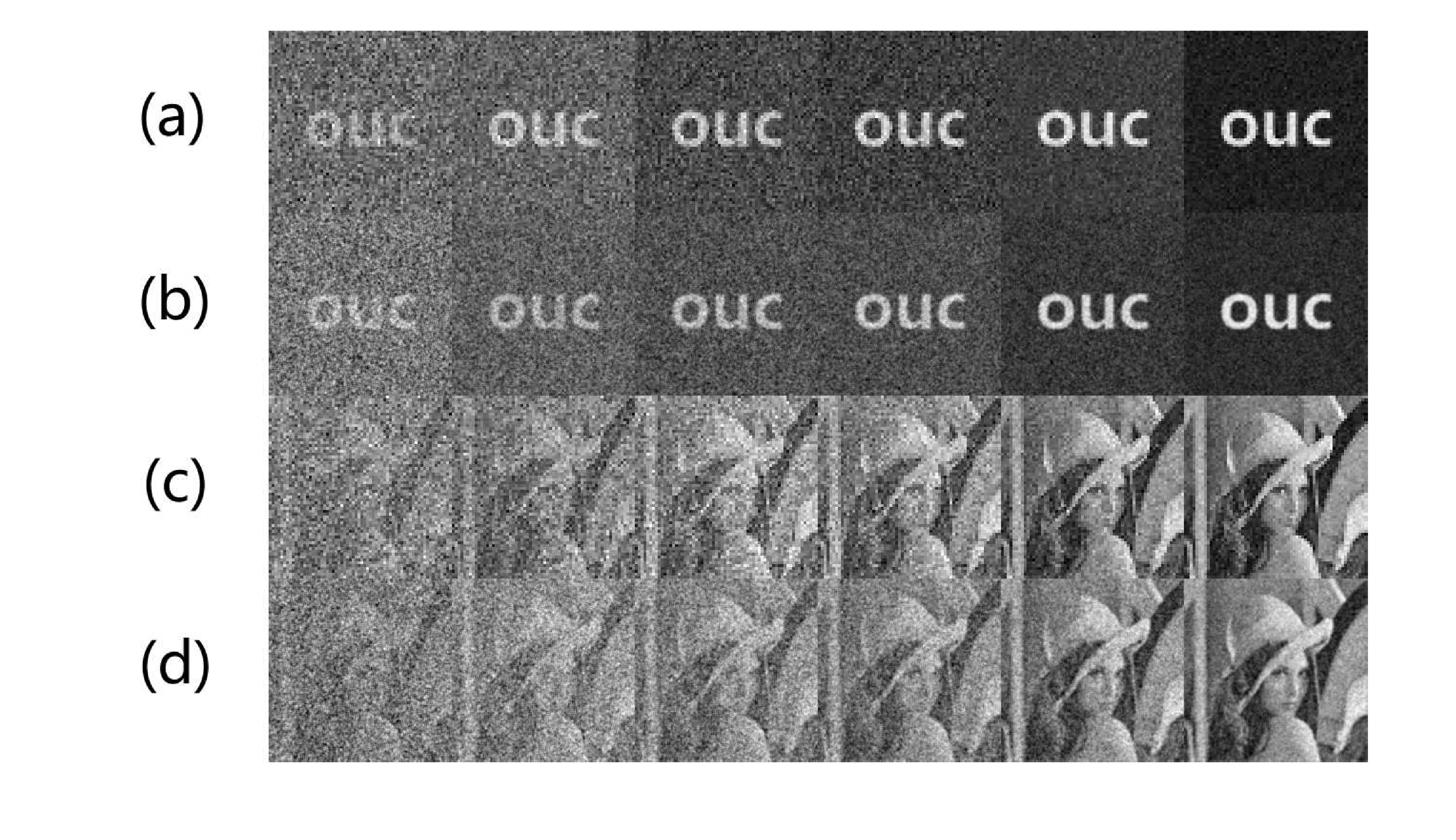}
\caption{Comparisons of the imaging quality of the two speckle patterns with size of 3*3, (a) and (c) are the results of random speckle patterns, (b) and (d) are the results of displacement speckle patterns, and the number of reconstructions is 1000, 3000, 6000, 10000, 30000, 80000 from left to right.}
\end{figure}
\begin{figure}[H]
\centering\includegraphics[width=7cm]{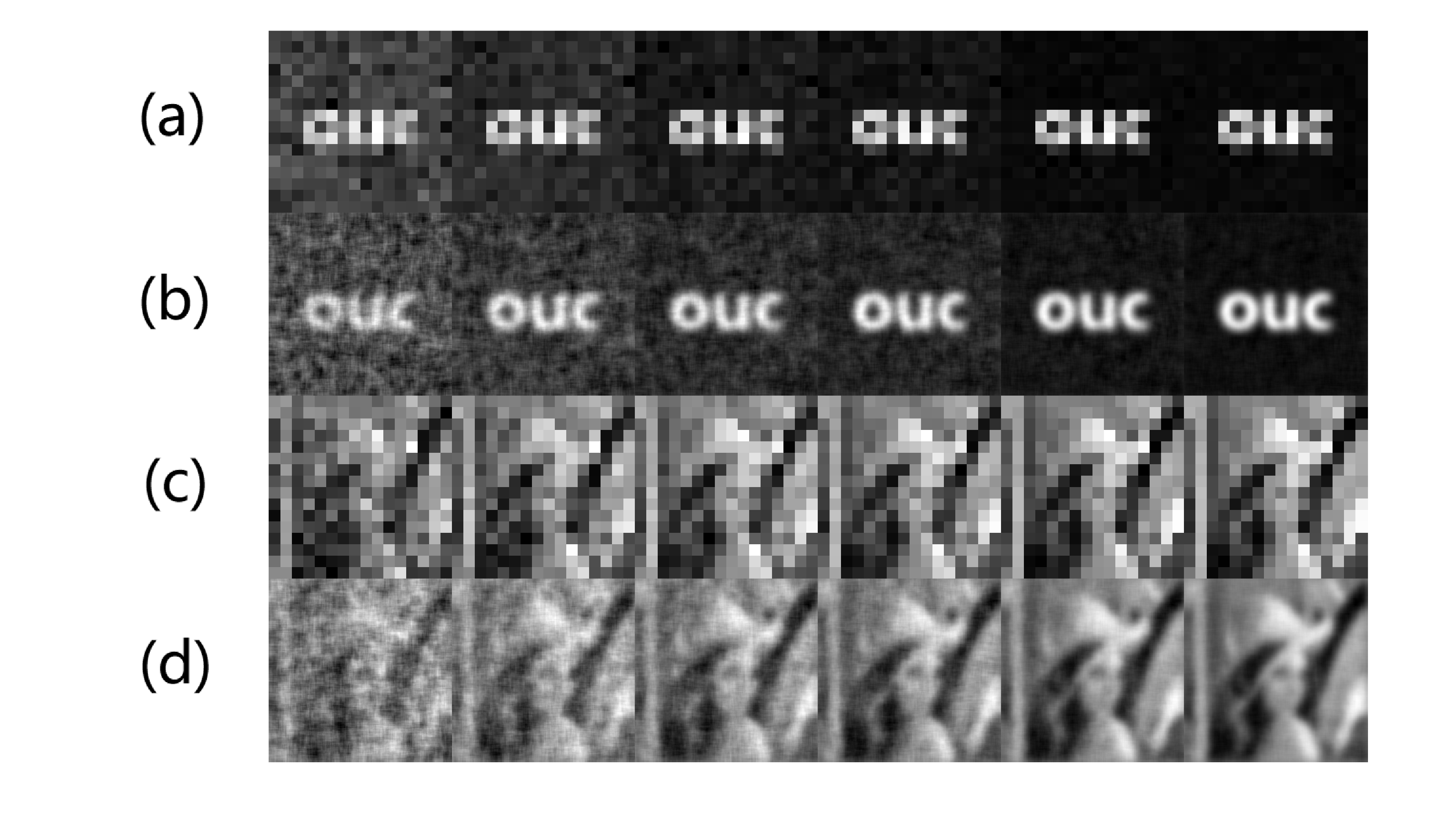}
\caption{Comparisons of the imaging quality of the two speckle patterns with size of 12*12, (a) and (c) are the results of random speckle patterns, (b) and (d) are the results of displacement speckle patterns, and the number of reconstructions is 1000, 3000, 6000, 10000, 30000, 80000 from left to right.}
\end{figure}

\par The PSNR curves with varying speckle sizes are shown in Fig.10. The red and black curves signify the PSNR of image reconstructed by the the displacement speckle patterns and random speckle patterns, respectively. From Fig. 10, we can see that, for the random speckle patterns, when the speckle size is taken between 8*8 and 12*12, the ghost image has the highest PSNR, but the more specific optimal speckle size cannot be obtained,because the speckle size cannot be taken continuously. For the displacement speckle patterns, we can know that when the speckle size is 10*10, the imaging quality is the best. This seems to show that compared with the random speckle patterns, the displacement speckle patterns can determine the optimal speckle size.
\begin{figure}[H]
\centering\includegraphics[width=7cm]{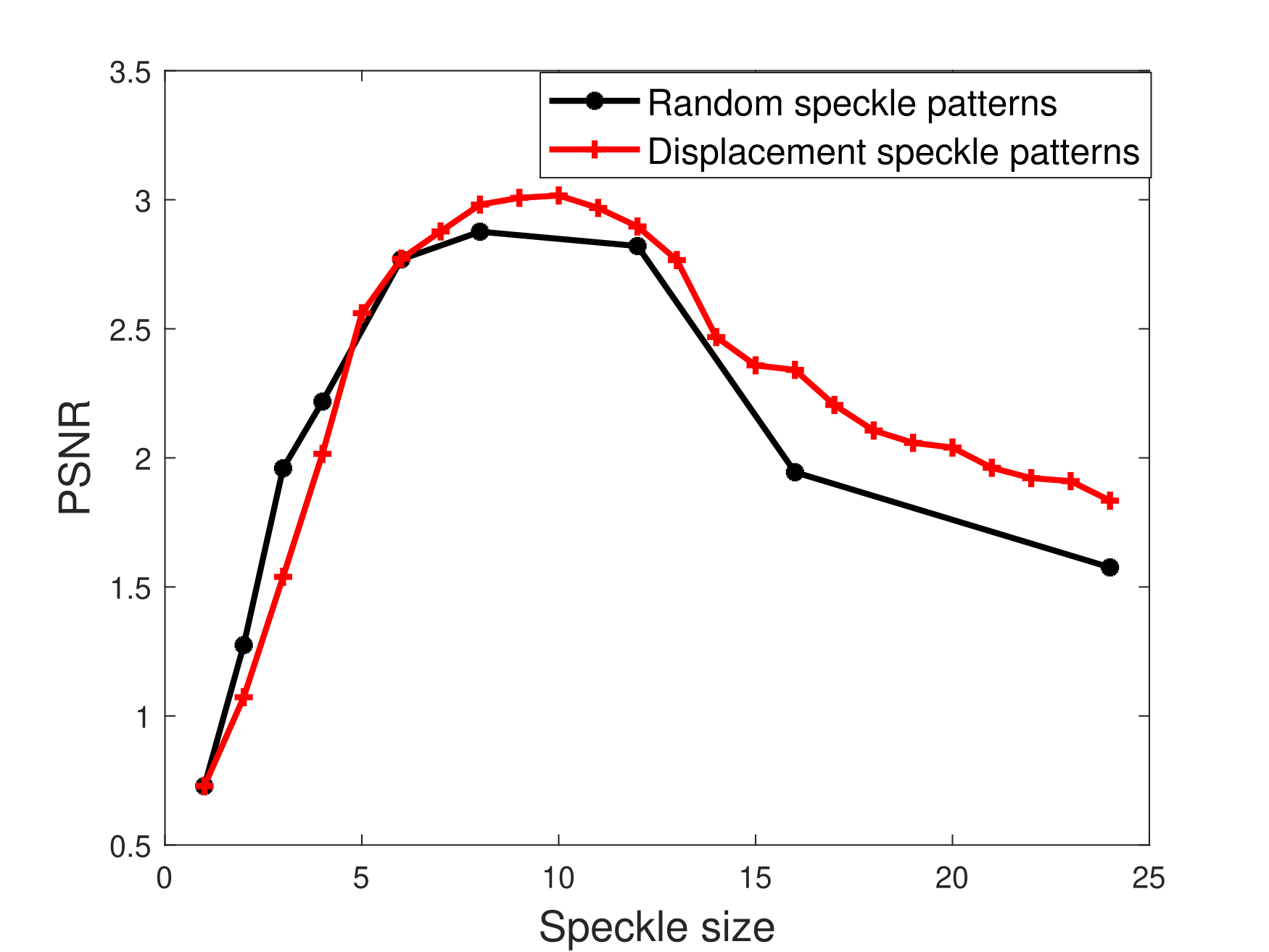}
\caption{The PSNRs as a function of the speckle size. The red curve is the PSNR of image reconstructed by the displacement speckle patterns;The black curve is the PSNR of image reconstructed by the random speckle patterns.}
\end{figure}

\section{Experimental verification of the effect of speckle patterns on ghost imaging}
\par According to the principle of computational ghost imaging, the experimental setup is built, as shown in Fig.11. The light source is a laser with a wavelength of 488nm. The light beam is  expanded by a beam expander and illuminate the DMD where the pre-designed speckle patterns are loaded. The light with speckle patterns is then projected to the surface of the object after passing through the lens filter system, and the reflected light is  collected by the bucket detector. Finally, the target image is obtained by executing a correlated calculation between the total intensity of light reflected by the object and the speckle patterns loaded on the DMD.
\begin{figure}[H]
\centering\includegraphics[width=10cm]{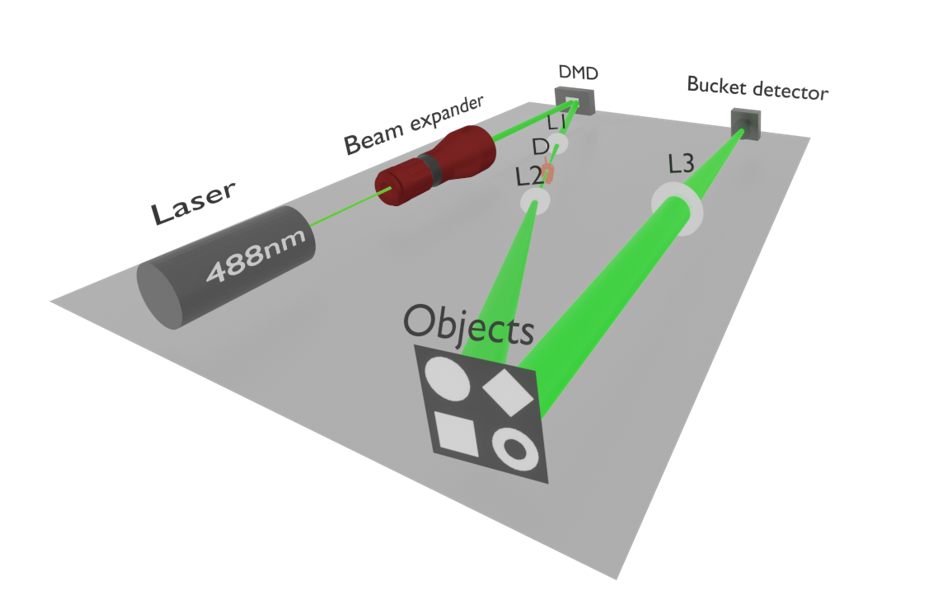}
\caption{The diagram of the experimental setup.}
\end{figure}
\par In order to verify the simulation conclusions in Section 3, this experiment uses the image composed of four shapes as the object to be measured, and selects the random speckle patterns and the displacement speckle patterns to compare the results. The speckle patterns consist of ten different sizes of speckles, including 6*6, 8*8, 12*12, 16*16, 19*19, 24*24, 38*38, 48*48, 57*57, 76*76, and the number of reconstructions is 4000. The part of experimental results are shown in Fig.12.

\begin{figure}[H]
\centering
\subfigure[]{
\includegraphics[width=1.5cm]{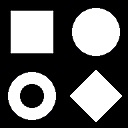}
%\caption{fig1}
}
\quad
\subfigure[]{
\includegraphics[width=5.5cm]{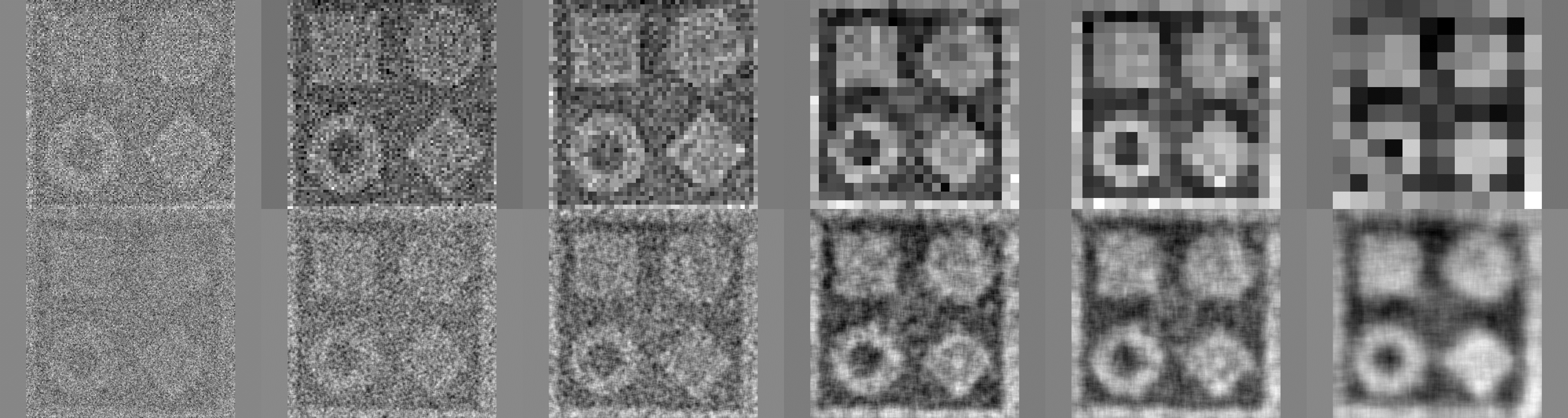}
}
\caption{(a)Objects. (b)The experimental results of random speckle patterns and displacement speckle patterns. The results shown in the top (bottom) line are  obtained with random (displacement) speckle patterns. The speckle sizes from left to right are 6*6, 12*12, 19*19, 38*38, 48*48, 76*76, respectively.}
\end{figure}
\par From Fig.12, it is obvious to see that when the speckle sizes are larger, the imaging qualities of the displacement speckle patterns are significantly improved compared to the random speckle patterns. To evaluate the imaging qualities of the two speckle patterns more objectively, we compare the PSNR of the retrived images. The PSNR curves of the two types of speckle patterns are shown in Fig.13:
\begin{figure}[H]
\centering\includegraphics[width=7cm]{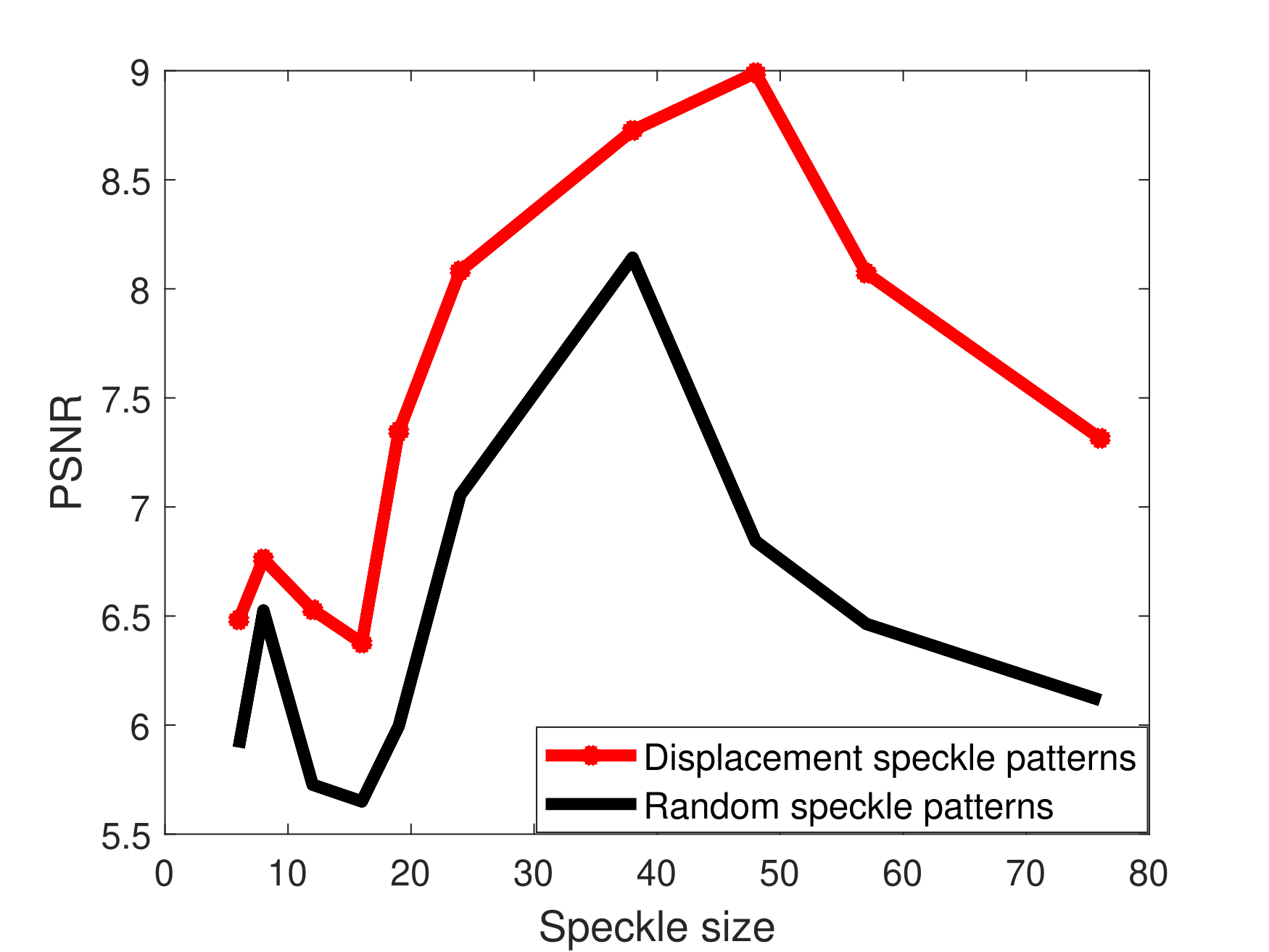}
\caption{The PSNR curves of the two types of speckle patterns. The red (black) curve is the PSNR of image reconstructed by the displacement (random) speckle patterns.}
\end{figure}
\par It can be seen from Fig.13 that the PSNR will improve with the increased speckle sizes from 16 to 48, and reachs the maxmium when the speckle size is 48, and then the PSNR will decrease.Due to the experimental errors, the PSNR fluctuates somewhat when the speckle size is less than 20. Moreover, the imaging qualities of the displacement speckle patterns are better than that of the random speckle patterns. The main experimental results are consistent with the simulation results.

\section{Conclusion}
%\section{Backmatter}
This paper explores the relationships between the ghost imaging qualities and the characters of speckle patterns, including the sizes, positions of speckles. We conclude that with the increase of the speckle size, the imaging quality can emerge optimal condition, that is, for a specific object, the speckle patterns with appropriate speckle size can be designed to improve the quality of ghost image. In addition, we proposed an optimized speckle patterns (displacement speckle patterns) in view of the shortage of the random speckle patterns. Through the comparative experiments, it is demonstrated that the proposed speckle patterns are more efficient for ghost imaging than the random speckle patterns when the speckle size is larger. So the displacement speckle patterns proposed can be used to improve the imaging qualities and reduce the number of reconstructions, which is of great significances for promoting the practical applications of CGI.

\par \textbf{Funding.}{This work was supported by the National Natural Science Foundation of China (Grants No. 61701464 and No. 61575180 ) and the Fundamental Research Funds for the Central Universities (Grants No. 202165008).}
\par \textbf{Acknowledgment.}The authors thank Xuan Chen and Xiaobing Hei for meaningful advice and discussions.
\par \textbf{Disclosures.} The authors declare no conflicts of interest.
\par \textbf{Data availability.} Data underlying the results presented in this paper are not publicly available at this time but may be obtained from the authors upon reasonable request.

\bibliography{sample}

\begin{thebibliography}{10}
\newcommand{\enquote}[1]{``#1''}

\bibitem{bina_backscattering_2013}
M.~Bina, D.~Magatti, M.~Molteni, A.~Gatti, L.~A. Lugiato, and F.~Ferri,
  \enquote{Backscattering differential ghost imaging in turbid media,}
  {\protect\JournalTitle{Phys. Rev. Lett.}} \textbf{110}, 083901 (2013).

\bibitem{gong_correlated_2011}
W.~Gong and S.~Han, \enquote{Correlated imaging in scattering media,}
  {\protect\JournalTitle{Opt. Lett.}} \textbf{36}, 394--396 (2011).

\bibitem{li_compressive_2020}
F.~Li, M.~Zhao, Z.~Tian, F.~Willomitzer, and O.~Cossairt, \enquote{Compressive
  ghost imaging through scattering media with deep learning,}
  {\protect\JournalTitle{Opt. Express}} \textbf{28}, 17395--17408 (2020).

\bibitem{yang_increasing_2015}
X.~Yang, Y.~Zhang, L.~Xu, C.-H. Yang, Q.~Wang, Y.-H. Liu, and Y.~Zhao,
  \enquote{Increasing the range accuracy of three-dimensional ghost imaging
  ladar using optimum slicing number method,} {\protect\JournalTitle{Chinese
  Phys. B}} \textbf{24}, 124202 (2015).

\bibitem{gong2016three}
W.~Gong, C.~Zhao, H.~Yu, M.~Chen, W.~Xu, and S.~Han, \enquote{Three-dimensional
  ghost imaging lidar via sparsity constraint,}
  {\protect\JournalTitle{Scientific reports}} \textbf{6}, 1--6 (2016).

\bibitem{erkmen_computational_2012}
B.~I. Erkmen, \enquote{Computational ghost imaging for remote sensing,}
  {\protect\JournalTitle{J. Opt. Soc. Am. A}} \textbf{29}, 782--789 (2012).

\bibitem{tong_preconditioned_2021}
Z.~Tong, Z.~Liu, C.~Hu, J.~Wang, and S.~Han, \enquote{Preconditioned
  deconvolution method for high-resolution ghost imaging,}
  {\protect\JournalTitle{Photon. Res.}} \textbf{9}, 1069--1077 (2021).

\bibitem{pittman_optical_1995}
T.~B. Pittman, Y.~H. Shih, D.~V. Strekalov, and A.~V. Sergienko,
  \enquote{Optical imaging by means of two-photon quantum entanglement,}
  {\protect\JournalTitle{Phys. Rev. A}} \textbf{52}, R3429--R3432 (1995).

\bibitem{zhang_correlated_2005}
D.~Zhang, Y.-H. Zhai, L.-A. Wu, and X.-H. Chen, \enquote{Correlated two-photon
  imaging with true thermal light,} {\protect\JournalTitle{Opt. Lett.}}
  \textbf{30}, 2354--2356 (2005).

\bibitem{chen_lensless_2009}
X.-H. Chen, Q.~Liu, K.-H. Luo, and L.-A. Wu, \enquote{Lensless ghost imaging
  with true thermal light,} {\protect\JournalTitle{Opt. Lett.}} \textbf{34},
  695--697 (2009).

\bibitem{shapiro_computational_2008}
J.~H. Shapiro, \enquote{Computational ghost imaging,}
  {\protect\JournalTitle{Phys. Rev. A}} \textbf{78}, 061802 (2008).

\bibitem{bromberg_ghost_2009}
Y.~Bromberg, O.~Katz, and Y.~Silberberg, \enquote{Ghost imaging with a single
  detector,} {\protect\JournalTitle{Phys. Rev. A}} \textbf{79}, 053840 (2009).

\bibitem{chan_high-order_nodate}
K.~W.~C. Chan, M.~N. O'Sullivan, and R.~W. Boyd, \enquote{High-order thermal
  ghost imaging,} {\protect\JournalTitle{Optics letters}} \textbf{34},
  3343--3345 (2009).

\bibitem{__2015}
K.~Qingnan, W.~Shande, Z.~Chi, and Q.~Weiping, \enquote{Influence of laser
  speckle average size on ghost imaging,} {\protect\JournalTitle{Optics and
  Precision Engineering}} \textbf{23}, 198--204 (2015).

\bibitem{chen_application_2014}
C.~Mingliang, L.~Enrong, and H.~Shensheng, \enquote{Application of
  multi-correlation-scale measurement matrices in ghost imaging via sparsity
  constraints,} {\protect\JournalTitle{Appl. Opt.}} \textbf{53}, 2924--2928
  (2014).

\bibitem{wang_effective_2020}
X.~Wang, Y.~Tao, F.~Yang, and Y.~Zhang, \enquote{An effective compressive
  computational ghost imaging with hybrid speckle pattern,}
  {\protect\JournalTitle{Optics Communications.}} \textbf{454}, 124470 (2020).

\bibitem{zhou_cheng_hybrid_2016}
Z.~Cheng, H.~Heyan, L.~Bing, and S.~Lijun, \enquote{Hybrid speckle-pattern
  compressive computational ghost imaging,} {\protect\JournalTitle{ACTA OPTICA
  SINICA.}} \textbf{36}, 0911001 (2016).

\bibitem{mao_speckle-shifting_2016}
T.~Mao, Q.~Chen, W.~He, Y.~Zou, H.~Dai, and G.~Gu, \enquote{Speckle-{Shifting}
  {Ghost} {Imaging},} {\protect\JournalTitle{IEEE Photonics J.}} \textbf{8},
  1--10 (2016).

\bibitem{nie_noise-robust_2021}
X.~Nie, F.~Yang, X.~Liu, X.~Zhao, R.~Nessler, T.~Peng, M.~S. Zubairy, and M.~O.
  Scully, \enquote{Noise-robust computational ghost imaging with pink noise
  speckle patterns,} {\protect\JournalTitle{Phys. Rev. A.}} \textbf{104},
  013513 (2021).

\bibitem{li2021sub}
Z.~Li, X.~Nie, F.~Yang, X.~Liu, D.~Liu, X.~Dong, X.~Zhao, T.~Peng, M.~S.
  Zubairy, and M.~O. Scully, \enquote{Sub-rayleigh second-order correlation
  imaging using spatially distributive colored noise speckle patterns,}
  {\protect\JournalTitle{Optics Express}} \textbf{29}, 19621--19630 (2021).

\bibitem{zerom_thermal_2012}
P.~Zerom, Z.~Shi, M.~N. O'Sullivan, K.~W.~C. Chan, M.~Krogstad, J.~H. Shapiro,
  and R.~W. Boyd, \enquote{Thermal ghost imaging with averaged speckle
  patterns,} {\protect\JournalTitle{Phys. Rev. A.}} \textbf{86}, 063817 (2012).

\bibitem{xuyang_xu_optimization_2015}
X.~X. Xuyang~Xu, E.~L. Enrong~Li, X.~S. Xia~Shen, and S.~H. Shensheng~Han,
  \enquote{Optimization of speckle patterns in ghost imaging via sparse
  constraints by mutual coherence minimization,} {\protect\JournalTitle{Chin.
  Opt. Lett.}} \textbf{13}, 071101 (2015).

\bibitem{wang_compressed_2020}
L.~Wang and S.~Zhao, \enquote{Compressed ghost imaging based on differential
  speckle patterns,} {\protect\JournalTitle{Chinese Phys. B.}} \textbf{29},
  024204 (2020).

\bibitem{cai_hongji_reflection_2019}
C.~Hongji, Y.~Zhihai, G.~Chao, R.~Jie, L.~Jiyuan, and W.~Xiaoqian,
  \enquote{Reflection ghost imaging based on superimposed speckle-pattern,}
  {\protect\JournalTitle{Laser \& Optoelectronics Progress}} \textbf{56},
  071101 (2019).

\bibitem{zhang2014object}
C.~Zhang, S.~Guo, J.~Cao, J.~Guan, and F.~Gao, \enquote{Object reconstitution
  using pseudo-inverse for ghost imaging,} {\protect\JournalTitle{Optics
  express}} \textbf{22}, 30063--30073 (2014).

\end{thebibliography}

\end{document}